\documentclass[preprint,showpacs,preprintnumbers,amsmath,amssymb,nofootinbib]{revtex4-1}

\usepackage{graphicx}% Include figure files
\usepackage{dcolumn}% Align table columns on decimal point
\usepackage{bm}% bold math

\newcommand{\beq}{\begin{equation}}
\newcommand{\eeq}{\end{equation}}
\newcommand{\beqy}{\begin{eqnarray}}
\newcommand{\eeqy}{\end{eqnarray}}
\newcommand{\beqyn}{\begin{eqnarray*}}
\newcommand{\eeqyn}{\end{eqnarray*}}
\newcommand{\nl}{\newline}

\newcommand{\mc}{\mathcal}

\newcommand{\slas}[1]{\not\!{#1}}
\newcommand{\bc}{\begin{center}}
\newcommand{\ec}{\end{center}}
\newcommand{\bmin}{\begin{minipage}}
\newcommand{\emin}{\end{minipage}}

\begin{document}

%\preprint{APS/123-QED}

\title{The problem of kinematic mass corrections  for unpolarized semi-inclusive deep inelastic scattering}

\author{Ekaterina Christova}
\email{echristo@inrne.bas.bg}

\author{Elliot Leader}

 \email{e.leader@imperial.ac.uk}
   \affiliation{Institute for Nuclear Research and Nuclear Energy \\ Sofia, Bulgaria}
\affiliation{Blackett laboratory \\Imperial College London \\ London SW7 2AZ, UK}

\date{\today}% It is always \today, today,

\begin{abstract}
Miraculously, target mass corrections for \emph{inclusive} deep inelastic scattering can be calculated exactly. On the contrary, there does not exist a consistent derivation of
 kinematic hadron  mass corrections for \emph{semi-inclusive} deep inelastic scattering (SIDIS). Recently this has become of topical interest, since there is a significant difference between the measured HERMES and COMPASS pion and kaon multiplicities, which cannot be explained as a consequence of evolution in $Q^2$, and it has been suggested that the difference can be understood if kinematic hadron mass corrections are taken into account. We explain why this argument is incorrect.
\end{abstract}

\maketitle

\section{\label{Int} Introduction}

Historically, the derivations of target-mass corrections  (TMC) for inclusive deep inelastic scattering (DIS) were all based on the
operator product expansion (OPE). The results for \emph{unpolarized}
 DIS were first derived by Nachtmann
\cite{Nachtmann:1973mr} employing a very elegant mathematical
approach in which the power series expansion used in the OPE was
replaced by an expansion into a series of hyperspherical functions
(representation functions of the homogeneous Lorentz group). Later,
also within the context of the OPE, Georgi and Politzer
\cite{Georgi:1976ve} re-derived Nachtmann's results using what they
called an alternative analysis ``for simple-minded souls like
ourselves" i.e.~based on a straightforward power series expansion
but, in fact, requiring a very clever handling of the combinatoric
aspects of the problem.

The derivation of target-mass corrections for \textit{polarized} DIS
turned out to be much more difficult. Several papers
\cite{Matsuda:1979ad,Wandzura:1977ce} succeeded in expressing the
reduced matrix elements $a_n$, $d_n$ of the relevant operators in
terms of combinations of moments of the structure functions, but did
not manage to derive closed expressions for the structure functions
$g_{1,2}$ themselves. The latter was finally achieved in 1997 by
Piccione and Ridolfi \cite{Piccione:1997zh} and later generalized to
weak interaction, charged current reactions, by Bl\"{u}mlein and
Tkabladze \cite{Blumlein:1998nv}.

Semi-inclusive deep inelastic scattering reactions, where a final-state hadron is monitored,  are of great interest, since they allow the extraction of information about individual antiquark distributions, and there is a major experimental effort underway to study them. However much of the most accurate data is, and will be for the forseeable future, in the kinematic region of relatively low $Q^2$, and it is thus of importance to know the kinematic hadron mass corrections (HMC) resulting from taking into account the target mass and produced hadron mass in these reactions.

The problem faced in deriving HMC for \emph{semi-inclusive} deep inelastic scattering (SIDIS) is that the OPE does not apply. For this reason, D'Alesio, Leader and Murgia searched for a method \emph{which does not rely on use of the OPE} and showed how the \emph{exact} TMC for DIS, both unpolarized and polarized, could be derived in a totally different approach \cite{D'Alesio:2009kv}.
 They  made the crucial observation that TMC, by definition, are \emph{kinematic} corrections, and therefore cannot depend on the \emph{numerical} value of the strong interaction coupling $g$. Thus they can be calculated exactly with $g=0$ i.e. using the ``handbag" diagram  as shown in Fig.~1. \nl
 \begin{figure}
 \includegraphics[width=0.5\textwidth]{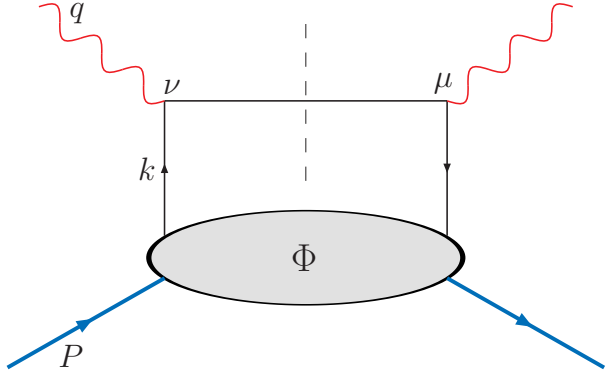}
 \caption{\label{Hand}The DIS ``handbag" diagram involving the $qq$-correlator.}
 \end{figure}

Christova and Leader (CL) thus attempted to apply this approach to
 calculate the exact HMC, to order $1/Q^2$, for unpolarized SIDIS \cite{ELKC}. Unfortunately they  found that there are serious problems and that
   the results break gauge invariance at the level of $(mass)^2/Q^2$.\footnote{For this reason we did not attempt to publish our work of 2011.}
    Moreover, as will be explained, it seems clear that this problem is not linked to the use of $g=0$ and is of a more fundamental nature.

   Recently Guerrero, Ethier, Accardi, Casper and Melnitchouk (GEACM) \cite{Guerrero:2015wha} presented a derivation
   of HMC for SIDIS and suggested that taking into account the HMC reduces, to a large extent, the difference between
   the HERMES and COMPASS pion and kaon multiplicities \cite{G}. Unfortunately, as we shall show,  the  GEACM derivation is inconsistent.

%%%%%%%%%%%%%%%%%%%%%%%%%%%%%%%%%%%%%%%%%%%%%%%%%%%%%%%%%%%%%%%%%%%%%%%%%%%%%%%%%%%%%%%%%%%%%%%%%%%%%%%%%%%%%%%
\section{\label{Not} Notation and conventions}

We shall largely follow the conventions of the classic paper (LM) of Levelt and Mulders  \cite{Levelt:1993ac}. We consider the SIDIS reaction
\beq\label{reaction}
 e(k_e) + N(P)\rightarrow e (k'_e) + h(P_h) + X
\eeq
where $N$ is the nucleon of mass $M$, $h$ is the detected hadron of mass $M_h$ and $X$ is the remainder of the final state.
 We use the standard DIS variables with $E$ and $E'$ the initial and final lepton energies in the target rest frame.
\beq \label{not}
 Q^2 = -q^2 \qquad \nu =\frac{P \cdot q}{M}= E-E' \qquad x_B=\frac{Q^2}{2M\nu} \qquad y= \frac{P\cdot q}{P\cdot k_e}= \frac{\nu}{E}
  \eeq
and the usual fragmentation variable $z_h$  defined as
\beq \label{olddef}
 z_h= \frac{P\cdot P_h}{P\cdot q} = \frac{E_h}{\nu}.
 \eeq
where $E_h$ is the energy of the produced hadron in the target rest
fame. The hadronic tensor for \emph{inclusive} DIS is denoted by
$W^{\mu\nu}$, and for \emph{semi-inclusive} DIS by
$\mathcal{W}^{\mu\nu}_h$. The particle label $h$ will occasionally
be left out for typographical clarity.

The unpolarized SIDIS cross-section is given by
\beq \label{Xsec}
\frac{2E_h\,d\sigma}{d^3 P_h dx_B dy } =\frac{\pi \alpha^2 \,y}{Q^4}\, L_{\mu\nu} \mathcal{W}^{\mu\nu}_h
\eeq
and the spin-averaged leptonic tensor is
\beq \label{LepT}
L^{\mu\nu} = 2 \, k_e^\mu \, k^{'\nu}_e + 2 \, k_e^\nu \, k^{'\mu}_e- Q^2 \, g^{\mu\nu}.
\eeq

%%%%%%%%%%%%%%%%%%%%%%%%%%%%%%%%%%%%%%%%%%%%%%%%%%%%%%%%%%%%%%%%%%%%%%%%%%%%%%%%%%%%%%%%%%%%%%%%%%%%%%%%%%%%%%%%%%%%%%%%%%%%%
\section{Expression for SIDIS hadronic tensor $\mc{W}^{\mu\nu}_h$ in terms of quark correlators}

\begin{figure}
 \includegraphics[width=0.7\textwidth]{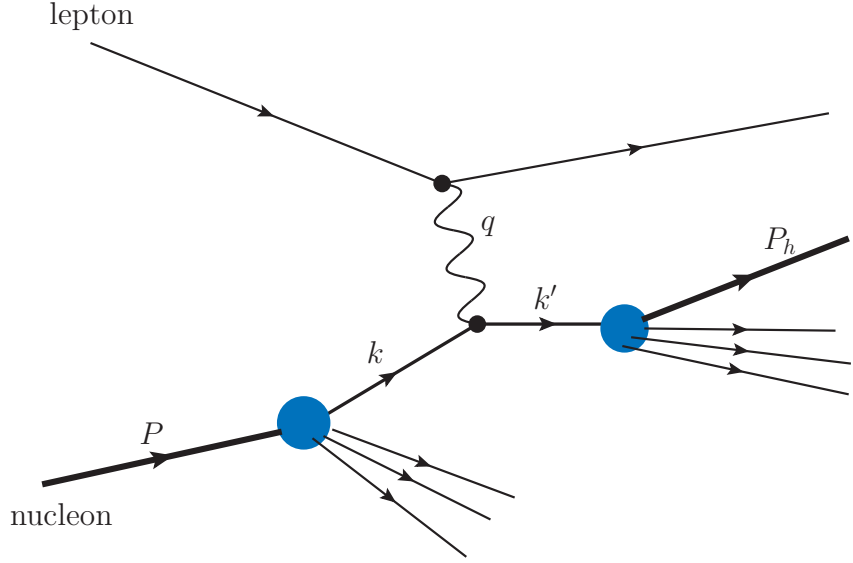}
 \caption{\label{Hand}The conventional partonic diagram for semi-inclusive lepton-nucleon reactions.}
 \end{figure}

From Fig.~2, for a quark of charge $e_q$ , for the
\emph{unpolarized} case we have: \beq \label{Wcalc}
\mc{W}_{un}^{\mu\nu}(P, P^h, q)= e_q^2 \, \int d^4k \, d^4 k'
\,\delta^4(k+q- k')\,Tr\left[ \gamma^\mu \Phi_q \gamma^\nu
\Delta_q^h \right], \eeq where $\Phi^q_{ij}(P,k)$ is the
spin-independent quark production correlator, with $k^\mu$ the
4-momentum of the active quark,  and $i,j$ Dirac indices, and
$\Delta_q^h(P_h, k')$ is the spin-independent quark fragmentation
correlator,
 with  $k'= k + q$  the momentum of the fragmenting  quark.  It is important, as will become clear presently, to keep separate the virtualities of the  the active quark and the fragmenting quark. We shall label these virtualities $m_q^2$ and $m_q^{'2}$
i.e. we take
\beq
 k^2\equiv (m_q)^2 \quad \textrm{and} \quad k'^2\equiv(m_q')^2.
  \eeq

In the usual treatment, where all hadron masses are ignored,
one takes $m_q=m'_q=0$, and  finds, that the leading twist expression for the SIDIS differential cross-section,  takes the form, for each flavour,
\beq \label{usualXsect}
\frac{d\sigma}{ dx_B dy dz_h}\propto q(x_B) D_q^h(z_h)
\eeq
where $q(x_B)$ is the usual quark-parton density (PDF) and  $D_q^h(z_h)$ the standard fragmentation function (FF).

 This result follows upon utilizing the leading twist expressions, which we shall refer to as ``order 1" i.e. $O(1)$, namely
\beq \label{PhiLT}
\Phi_q(x_B)\equiv \int dk^- \, d^2\bm{k}_\bot \Phi_q(P,k) = \frac{1}{2}q(x_B) \slas{\bar{n}}
 \eeq
\beq \label{ChiLT}
 \Delta_q^h(z_h)\equiv \int dk^{'+} \, d^2\bm{k'}_\bot \Delta_q^h(P_h, k')= \frac{1}{2}D_q^h(z_h)\slas{n},
  \eeq
where the GEACM null vectors are defined as\footnote{These null vectors are  almost universally labeled $n_+$ and $n_-$ in the literature.}
\beq
 \bar{n}^\mu=\frac{1}{\sqrt{2}}(1,0,0,1) \qquad \textrm{and}\qquad n^\mu=\frac{1}{\sqrt{2}}(1,0,0,-1),
 \eeq
and neglecting terms of $O(\mc{M}/Q)$ and  $O(\mc{M}^2/Q^2)$, where $\mc{M}^2$ generically stands for $M^2$,\,$M_h^2$ or $MM_h$,
 when evaluating $\mc{W}_{un}^{\mu\nu}(P, P^h, q)$ in Eq.~(\ref{Wcalc}).

%%%%%%%%%%%%%%%%%%%%%%%%%%%%%%%%%%%%%%%%%%%%%%%%%%%%%%%%%%%%%%%%%%%%%%%%%%%%%%%%%%%%%%%%%%%%%%%%%%%%%%%%%%%%%%%%%%
\section{The approach of GEACM and its problems}

In their treatment of the HMC, GEACM  utilize Eqs.~(\ref{PhiLT}, \ref{ChiLT}) in Eq.~(\ref{Wcalc}), but
assume collinear production i.e. put $\bm{P}^h_\bot = 0 $, arguing that the transverse momentum should be  generated by interactions, and then treat the kinematics in Eq.~(\ref{Wcalc}) more carefully, keeping all terms of $O(\mc{M}^2/Q^2)$.
Their key result is that  Eq.~(\ref{usualXsect})   is then replaced by
\beq \label{corXsec}\frac{d\sigma}{ dx_B dy dz_h}\propto q(\xi_h) D_q^h(\zeta_h), \eeq
where
\beq \label{Chi} \xi_h= \xi \left(1 + \frac{(m'_q)^2}{Q^2}\right) \eeq
and
\beq \label{Zeta}\zeta_h=\frac{z_h \xi}{2 x_B}\,\left( 1 +\sqrt{1 - \frac{4 x_B^2M^2M_h^2}{z_h^2 Q^4}}\right).
\eeq
Here $\xi$ is the usual Nachtmann variable:
\beq \label{Nach}
\xi=\frac{2x_B }{1 + \sqrt{1+4x_B^2 \, M^2 / Q^2}}.
\eeq

It is clear that the GEACM result differs from the conventional massless result by terms of order
 $O(\mc{M}^2/Q^2)$. Thus to be consistent and believable the  GEACM evaluation of $\mc{W}_{un}^{\mu\nu}(P, P^h, q)$ in Eq.~(\ref{Wcalc})
 must be correct to $O(\mc{M}^2/Q^2)$. Now $\mc{W}_{un}^{\mu\nu}(P, P^h, q) $ involves a product of $\Phi_q $ and $\Delta_q^h$
 so that to achieve the desired accuracy each of $\Phi_q $ and $\Delta_q^h$ must be given correct to $O(\mc{M}^2/Q^2)$.
  But this is not done! Eqs.~(\ref{PhiLT}, \ref{ChiLT}) are only correct to O(1). Hence the GEACM result is not consistent.

We shall now indicate the, what to us seem like insurmountable difficulties,
that arise if  we try to remedy this problem in a straightforward way.

Correct to $O(\mc{M}^2/Q^2)$ the quark production correlator involves 3 scalar functions and has the form:
\beq \label{Phi}
\Phi_q(x)= \frac{M}{2P_+}\, e(x) +\frac{q(x)}{2}\slas{\bar{n}} + \frac{M^2}{2(P_+)^2}\, b(x)\slas{n}
\eeq
where, in what GEACM call the Breit Frame, $P_+=O(Q)  $ .
It might be thought that the extra functions appearing in Eq.~(\ref{Phi})  are a consequence of interactions and therefore
can be ignored in  a purely kinematic analysis , but according to Mulders and Tangerman
\cite{Mulders:1995dh} this is incorrect. They show that  e.g.
\beq \label{e} e(x)=e_{kin}(x)  + e_{int}(x) ,      \eeq
where
\beq \label{ekin}
 e_{kin}(x)= \frac{m_q}{xM}  q(x)  .
  \eeq
A completely analogous     development holds for the  fragmentation correlator
 $\Delta_q^h$, which  then also   contains 3 terms, parts of which are definitely not due to interaction.

The most serious consequence of using     $\Phi_q(x)$ and $ \Delta_q^h$ ,    correct to $O(\mc{M}^2/Q^2)$ , is the breakdown of
 gauge    invariance for  $\mc{W}_{un}^{\mu\nu}(P, P^h, q) $, which we will now explain.

%%%%%%%%%%%%%%%%%%%%%%%%%%%%%%%%%%%%%%%%%%%%%%%%%%%%%%%%%%%%%%%%%%%%%%%%%%%%%%%%%%%%%%%%%%%%%%%%%%%%%%%%%%%%%%%%%%%%%%%%%%
 \section{The breakdown of gauge invariance: a simple demonstration}
 We are only interested in   kinematical corrections. A simple trick to isolate these is thus to switch off the strong
 interaction i.e. to take $\alpha_s=0$. Then, according to    \cite{D'Alesio:2009kv}, the  expressions for   the
 corrected correlators become
 \beq \label{Phi0} \Phi_q(x) \propto q(x)[\,m_q + \slas{k}\,] \eeq
 \beq \label{Chi0} \Delta_q^h(z)\propto D_q^h(z)[\,m'_q + \slas{k}'\,] \eeq
 which lead to
 \beq \label{W00}
  \mc{W}_{un}^{\mu\nu}\, \propto \,( m_qm'_q-k\cdot k')\, g^{\mu\nu} + (k^\mu k^{'\nu} + k^ \nu k^{'\mu})  .
   \eeq
   Gauge invariance requires  that
   \beq \label{GI}
   q_\mu W^{\mu\nu}_{un} =0.
   \eeq
   Using Eq.~(\ref{W00}) we find
   \beq \label{noGI}
     q_\mu W^{\mu\nu}_{un}\,\propto \,(m'_q - m_q)\, [\,m_q q^\nu + (m'_q + m_q) k^\nu\,]
     \eeq

In other words, gauge invariance demands that $m_q=m'_q$. Is this possible?

In the standard treatment, ignoring hadron masses,   one conventionally takes  $m_q=m'_q=0$ and gauge invariance is fine.
 When hadron masses are included there are compelling reasons to still choose   $m_q=0$, as GEACM do, but it is certainly
  incorrect to take
 $m'_q=0 $.     Indeed,  kinematical considerations imply a lower bound  for $ m^{'2}_q $.
   For the collinear case CL also have it:
 \beq \label{LB}
   m^{'2}_q \geq M_h^2/z_h
    \eeq
 which is compatible with the CEACM lower bound $ m^{'2}_q \geq M_h^2/\zeta_h $.
 \footnote{The GEACM result assumes that the target remnant jet has $(\textrm{mass})^2\equiv(P-k)^2\geq M^2$,
 an assumption which we don't think can be justified, given that the remnant jet has baryon number 2/3
  and is coloured, and which disagrees with the condition $(P-k)^2\geq 0$ used by Ellis, Furmanski and Petronzio \cite{Ellis:1982cd}.}
 \emph{This immediately implies that  we cannot take $m_q=m'_q$ and we are forced to conclude that when terms of} $O(\mc{M}^2/Q^2)$
 \emph{are included consistently in the} GEACM \emph{appraoach, the result is not gauge invariant.} \nl
  (In addition to this problem, there is another worrying matter. In their paper GEACM choose the particular value $M_h^2/\zeta_h$
   for $m^{'2}_q$. But any value larger than this would be acceptable. Hence there is effectively an arbitrary parameter in the GEACM treatment.)

 Consider now the implications of the lower bound in Eq.~(\ref{LB}).    As stressed by Mulders and Tangerman  \cite{Mulders:1995dh}
  the validity of the parton model  in QCD depends on the \emph{assumption} that all the quark   correlators
   cut off rapidly with increasing quark virtuality, implying that the fragmentation correlator should cut off rapidly
 with increasing $k^{'2}\equiv m_q^{'2}$.  But use of  Eq.~(\ref{LB}) to describe experimental multiplicity values which are not small for small
 values of    $z_h$ would imply that   the correlator is     large for virtualities much greater
  than a $(\textrm{hadrom mass})^2 $.

%%%%%%%%%%%%%%%%%%%%%%%%%%%%%%%%%%%%%%%%%%%%%%%%%%%%%%%%%%%%%%%%%%%%%%%%%%%%%%%%%%%%%%%%%%%%%%%%%%%%%%%%%%%%%%%%%%%%%%%%%%%%%

   \section{Possible resolutions of the problem??}

  The above discussion suggests that there is no way to accommodate non-zero hadron masses in the conventional treatment
  of SIDIS reactions.
  We list here, with some lack of conviction,  a couple of
    unconventional ways to overcome the difficulties.
    \begin{figure}
 \includegraphics[width=0.9\textwidth]{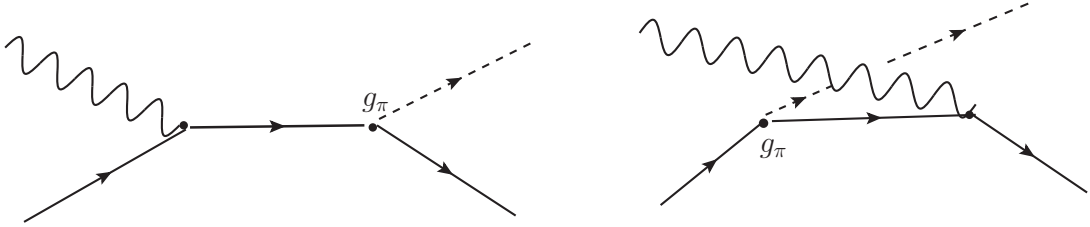}
 \caption{\label{Hand}Feynman diagrams for pion photoproduction on a quark. Only the sum of both diagrams is gauge invariant. The analogue of the crossed diagram is missing in Fig.~2.}
 \end{figure}
 \begin{enumerate}
 \item Since the virtuality of the fragmenting quark is considerably larger than   the square of a typical hadron mass, it  is neither a typical partonic quark nor a constituent quark. It is therefore some kind of effective quark and as such one might introduce an effective electromagnetic coupling  e.g.
        \beq
        \gamma ^\mu \rightarrow  \gamma^\mu-\frac{\slas{q}}{q^2}q^\mu.
        \eeq
    It would then be possible to achieve a gauge invariant result, while keeping $m'_q\neq m_q$ .
    \item By analogy with the treatment of $\pi ^0$-Photoproduction on a quark, one can  restore gauge
    invariance by including
    the crossed Feynman diagram shown in Fig.~3, in which, in the pion-quark coupling $g_\pi\gamma_5$, the    constant $g_\pi$ is replaced by a phenomenological scalar function. This was tried by CL \cite{ELKC}, but they  were unable to reproduce the standard result in the limit
    $Q^2\rightarrow  \infty $.
    \end{enumerate}

%%%%%%%%%%%%%%%%%%%%%%%%%%%%%%%%%%%%%%%%%%%%%%%%%%%%%%%%%%%%%%%%%%%%%%%%%%%%%%%%%%%%%%%%%%%%%%%%%%%%

\section{Conclusions} Guerrero, Ethier, Accardi, Casper and Melnitchouk have produced a study of semi-inclusive deep inelastic scattering, which attempts to take into account the masses of the target and produced hadron, contrary to the conventional treatment which ignores all hadronic masses.  They then argue that such effects might reduce the apparent discrepancy between
the HERMES and COMPASS pion and kaon multiplicity measurements.    \nl
Unfortunately it turns out that the GEACM analysis is inconsistent, in that terms of the same order of magnitude as those they are concerned about, are neglected. Moreover,  when such terms are included the resulting hadronic tensor is no longer gauge invariant. \nl
We have, regrettably, been forced to conclude, that in contradistinction to inclusive DIS, where it is possible to calculate exact target mass corrections, attempts to include kinematic hadron mass corrections in semi-inclusive DIS run into insurmountable difficulties. It seems that the standard formulation of a semi-inclusive event, as a product of  a parton density times an independent  fragmentation function does not work if hadron masses are taken into account.

\acknowledgements{E.L. is grateful to the Leverhulme Trust for an Emeritus Fellowship.

%\bibliography{Elliot_General}

\end{document}